\documentclass[prd, twocolumn, nofootinbib, floatfix]{revtex4-1} 

\usepackage[mathscr]{euscript}
\usepackage{amsmath}
\usepackage{graphicx}
\usepackage{dcolumn}
\usepackage{bm}
\usepackage{epsfig}
\usepackage{amssymb,latexsym,mathrsfs}
\usepackage{graphicx}
\usepackage{color}
\usepackage{hyperref}
\usepackage{float}
\usepackage{diagbox}
  
\definecolor{vividviolet}{rgb}{0.62, 0.0, 1.0}
\definecolor{amaranth}{rgb}{0.9, 0.17, 0.31}
\definecolor{palatinateblue}{rgb}{0.15, 0.23, 0.89}
\definecolor{brightpink}{rgb}{1.0, 0.0, 0.5}
\definecolor{cornflowerblue}{rgb}{0.39, 0.58, 0.93}
\definecolor{deepcarminepink}{rgb}{0.94, 0.19, 0.22}
\definecolor{radicalred}{rgb}{1.0, 0.21, 0.37}

\hypersetup{ linktoc=all,
    colorlinks, linkcolor={palatinateblue},
    citecolor={brightpink}, urlcolor={amaranth}
}

\newcommand{\be}{\begin{equation}}
\newcommand{\ee}{\end{equation}}
\newcommand{\bs}{\begin{split}} 
\newcommand{\bea}{\begin{eqnarray}}
\newcommand{\eea}{\end{eqnarray}}

\renewcommand{\d}[1]{\ensuremath{\operatorname{d}\!{#1}}}

\begin{document}

\title{Accelerating Boundary Analog of a Kerr Black Hole} 

\author{Michael R.R. Good${}^{1,2}$}
\email{michael.good@nu.edu.kz}
\author{Joshua Foo${}^{3}$}
\email{joshua.foo@uqconnect.edu.au}
\author{Eric V. Linder${}^{2,4}$}
\email{evlinder@lbl.gov}
\affiliation{${}^1$Physics Department, Nazarbayev University, 
Nur-Sultan, Kazakhstan\\
${}^2$Energetic Cosmos Laboratory, Nazarbayev University, Nur-Sultan, Kazakhstan\\ 
${}^3$Centre for Quantum Computation \& Communication Technology, School of Mathematics \& Physics,
The University of Queensland, St.~Lucia, Queensland, 4072, Australia\\ 
${}^4$Berkeley Center for Cosmological Physics \& Berkeley Lab, University of California, Berkeley, CA, USA  
}

\begin{abstract} 
An accelerated boundary correspondence (i.e.\ a flat spacetime accelerating mirror trajectory) is derived for the Kerr spacetime, with a general  formula that ranges from the Schwarzschild limit (zero angular momentum) to the extreme maximal spin case (yielding asymptotic uniform acceleration).  The beta Bogoliubov coefficients reveal the particle spectrum is a Planck distribution at late times with temperature cooler than a Schwarzschild black hole,  due to the `spring constant' analog of angular momentum. The quantum stress tensor indicates a constant emission of energy flux at late times consistent with eternal thermal equilibrium. 
\end{abstract} 

\date{\today} 

\maketitle 

\section{Introduction}\label{sec:Intro}
Essentially all astrophysical black holes observed in our universe are described by the Kerr metric. Quasars in active galactic nuclei, supermassive black holes in the centers of galaxies, and black holes in binary systems whose eventual inspiral can be measured by gravitational wave detectors possess angular momentum, and can be analyzed using the Kerr metric or its perturbations. Thus, a deeper understanding of the Kerr metric \cite{kerr1963gravitational} -- and how it affects quantum particle production from black hole event horizons -- is a pertinent question, highly relevant to the fields of quantum cosmology and quantum field theory in curved spacetime. However due to the nonlinearity of the metric, exact calculations in the Kerr spacetime are difficult and even intractable, in comparison to its non-rotating partner, the Schwarzschild metric. 

To investigate the quantum phenomena produced by Kerr black holes, we consider a cleaner analog system, namely an accelerating boundary in flat spacetime, which replaces gravitational effects with those induced by acceleration. A well-known prediction of relativistic quantum field theory is that accelerated mirrors radiate particles, a manifestation of the dynamical Casimir effect \cite{Birrell:1982ix,moore1970quantum}. Mirrors act as boundaries which act on and transform incoming field states, especially the vacuum, of quantum fields. For appropriately chosen trajectories, the radiation flux is thermal, and comparisons can be made with the Hawking radiation emitted from a black hole formed via gravitational collapse \cite{Hawking:1974sw,DeWitt:1975ys, Davies:1976hi,Davies:1977yv,walker1985particle,carlitz1987reflections,Su_2017}. Recent experiments have also been proposed to observe this effect; see for example \cite{Chen:2020sir, Chen:2015bcg,blencowe2020analogue,PhysRevA.99.053833}. 

While accelerated mirrors have been long-studied in the literature (see for example, \cite{yab,Ford:1982ct,Dodonov:2020eto}), recent works \cite{good2013time,Romualdo:2019eur,Good_2015BirthCry,fullingpage,Good:2018aer,Cong:2018vqx,Good:2017kjr,Fulling:2018lez} have revisited the problem and demonstrated that new insights can still be obtained. In particular, the application of the so-called accelerated boundary correspondence (ABC) has yielded exact results for the particle and energy spectrum of Hawking radiation emitted from nontrivial black hole spacetimes, along with its thermodynamical properties. Recent studies have analyzed the Schwarzschild \cite{Good_2017Reflections,Anderson_2017,Good_2017BHII}, Schwarzschild with Planck length \cite{good2020schwarzschild}, Reissner-Nordstr\"om (RN) \cite{good2020particle}, and extreme RN \cite{good2020extreme} cases through a transformation of the (3+1)-dimensional metric to a (1+1)-dimensional mirror trajectory in flat spacetime. This approach has also been applied to the cosmological horizon of de Sitter space, where an exact thermal distribution was derived \cite{Good:2020byh}. 

In this paper, we build on previous works and derive an ABC for the axially symmetric, rotating Kerr metric. 
We derive in Sec.~\ref{sec:metacc} the relation between the Kerr metric, null shell collapse, and the matching condition for the accelerated mirror trajectory. 
In Sec.~\ref{sec:particles}, we calculate the quantum particle spectrum and compare it to the late-time Schwarzschild mirror solution (the eternal black hole spectrum of the Carlitz-Willey mirror \cite{carlitz1987reflections}). We extend the mapping to the extremal Kerr spacetime (EK) in Sec.~\ref{sec:EK} and conclude in Sec.~\ref{sec:conc}.

\section{From Kerr Metric to Acceleration} \label{sec:metacc} 

\subsection{Kerr Metric} 

The Kerr metric is not spherically symmetric, but this is immaterial to the coordinates describing the collapsing star's center. The regularity condition at the origin defines the behaviour of incoming modes, subsequently governing the character of outgoing particle production \cite{wilczek1993quantum,Rothman:2000mm, Fabbri}. The position of the center of the star is insensitive to the asymmetry of the surface event horizon and 
a (1+1)-dimensional model can be constructed by specializing to a single plane. Hence, the angular coordinates are not required in the calculation of the temperature of the radiation. This was discovered for the late-time extremal Kerr case by Rothman \cite{Rothman:2000mm}. We will demonstrate this fact holds true for both the extremal and non-extremal Kerr spacetimes at all times. 

The line element in spatial spherical coordinates is 
(throughout this paper we utilize natural units, $G = h = c  = k_B = 1$) 
\begin{equation}
ds^{2}=g_{tt}dt^{2}+g_{rr}dr^{2}+g_{\theta\theta}d\theta
^{2}+g_{\phi\phi}d\phi^{2}+2g_{t\phi}dtd\phi\text{,}\label{kerrmetric}
\end{equation}
with%
\begin{align}
g_{tt}  &  =-\left(  \Delta-a^{2}\sin^{2}\theta\right)  \Sigma
^{-1},\text{ \ \ }g_{rr}=\Sigma\Delta^{-1},\text{ }\nonumber\\
\text{\ }g_{\theta\theta}  &  =\Sigma,\text{ \ }g_{\phi\phi}=-\sin^{2}%
\theta\left[  a^{2}\Delta\sin^{2}\theta-\left(  r^{2}+a^{2}\right)
^{2}\right]  \Sigma^{-1},\nonumber\\
\text{\ }g_{t\phi}  &  =-a\sin^{2}\theta\left(  r^{2}+a^{2}%
-\Delta\right)  \Sigma^{-1},
\end{align}
where%
\begin{align}
\Delta &  \equiv a^{2}+r^{2}-rr_{s},\text{ \ }\Sigma\equiv a^{2}%
\cos^{2}\theta+r^{2},\text{ \ }\nonumber\\
r_{s}  &  \equiv 2M,\text{ \ } a\equiv J/M\,.
\end{align} 
In this parametrization, $a$ is the `spin' or mass-normalized angular momentum of the rotating black hole and $r_s = 2M$ is the usual Schwarzschild radius. 

The corresponding static coordinate metric of the (1+1)-dimensional Kerr metric can be obtained by setting $\phi = \theta = 0$ in the (3+1)-dimensional case, Eq.~(\ref{kerrmetric}). This yields the following line element in terms of the radial and time pieces,
\be \label{dsKerr} \d s^2 = -f(r)\d t^2 + f(r)^{-1}\d r^2\,,  \ee 
with 
\be f(r)= 1 - \frac{2M r}{r^2 + a^2}\,.\ee
Note that this is the same $a$ as that found in Eq.~(\ref{kerrmetric}). The metric in (1+1) dimensions is introduced to find the associated radial trajectory (i.e.\ the mapping of inside to outside coordinates, see Sec.~\ref{sec:K}) of the center of the black hole in (3+1) dimensions, which can be understood as the reflecting point of the incoming modes. Hence, the mirror trajectory in (1+1) is an appropriate model for describing the collapse and its subsequent effect on the modes, as far as an analysis of the radiation temperature is concerned. The angular coordinates $\theta$ and $\phi$ do not affect the temperature since they do not enter the equation for $f(r)$. This will illustrate an important 1-dimensional trait of the Kerr black hole, similar to single channel Bekenstein entropy flow \cite{Bekenstein:2003dt, Padmanabhan:2002sha}.  The resulting temperature is for the equivalent  (3+1)-dimensional Kerr black hole. 

Equation~(\ref{dsKerr}) contains two horizons at the radial coordinates $r_{p,m} =M\pm \sqrt{M^2-a^2}$ (where $r_p,r_m$ correspond to the $\pm$ signs respectively), which reduce to the event horizon, $r = r_s$, and the curvature singularity, $r = 0$, of the Schwarzschild metric in the limit $a \to 0$. Using Eq.\ (\ref{dsKerr}) one can straightforwardly derive the analogous single moving mirror model  \cite{Davies:1976hi,Davies:1977yv} trajectory, where the accelerating boundary plays the role of the black hole center. The resulting particle production emitted by the mirror can be analyzed through the calculation of Bogoliubov coefficients between the incoming and outgoing modes.

\subsection{Accelerated Boundary Correspondence} \label{sec:K} 

In Kerr spacetime, the temperature of emitted radiation observed by an inertial observer at infinity is 
\be 
T_K = \frac{\kappa}{2\pi}=\frac{g}{2\pi} - \frac{k}{2\pi}=\frac{1}{8\pi M}\frac{2\beta}{1+\beta}\,,\label{eq:tk} 
\ee 
where $\beta=\sqrt{1-a^2/M^2}$ 
(see Appendix~\ref{appx} for a derivation). We will show that the same Planck spectrum with an identical temperature holds for the accelerating mirror analog. Here $g=1/(4M)$ is the usual Schwarzschild surface gravity, and $k=M\Omega^2$ is the black hole spring constant \cite{Good:2014uja}. The parameter $\beta$ 
will allow us to present results for the continuous range from 
Schwarzschild ($\beta=1$) to extreme Kerr ($\beta=0$) solutions. 

For a double null coordinate system $(u,v)$, with $u = t-r^*$ and $v = t+r^*$, the appropriate tortoise coordinate $r^*$ is found in the usual way, via
\be r^* = \int \mathrm{d} r \: f(r)^{-1} ,\label{tort}\ee 
yielding
\be r^* = r+ M \ln \left|\frac{(r-r_p)(r-r_m)}{r_s^2}\right|+\frac{M}{\beta}\ln \left|\frac{r-r_p}{r-r_m}\right|\,.  \label{eq:tort} 
\ee
One then has the metric for the geometry describing the outside region  $r>r_p$, 
\be \d s^2 =-f \; \d u \d v\,.\ee 

The matching condition (see e.g.\ \cite{wilczek1993quantum,Fabbri}) with the flat interior geometry, described by the interior coordinates $U=T-r$ and $V=T+r$, is the trajectory of $r=0$, expressed in terms of the exterior function $u(U)$ with interior coordinate $U$. 
This matching is obtained via the association, $r^* = r$.  We take $r^*(r=(v_0 -U)/2) = (v_0-u)/2$, occurring along a light ray, $v_0$.  We can choose either $v_0 - 2 r_p \equiv v_H$ or $v_0-2r_m\equiv v_H$ because $u\rightarrow +\infty$ at $U=v_H$. Without loss of generality, we can set $v_H= 0$, i.e.\ $v_0=2r_{p,m}$. 
Choosing $v_0=2r_p$ gives the correct Schwarzschild limit since for Schwarzschild, $r_m=0$. Another way of understanding this choice is that the modes that escape the incipient black hole necessarily have access to the $r=0$ center. Anticipating a transition to the mirror system, the outer radius is chosen for the shell position because $r_p>r_m$. That is, the modes from the shell $v_0 = 2r_p$ will reach the observer at $\mathscr{I}_R^+$ first in both the mirror and black hole system, already having passed through $r=0$ (reflecting off the mirror). Thus, we choose $v_0 = 2r_p$ rather than the inner radius which occurs at an earlier $v$. 

Taking the outer horizon $v_0=2r_p$, the result for the exterior coordinate is expressed as  
\be 
u(U)=U-2M\frac{1+\beta}{\beta}\ln\left|\frac{U}{4M}\right|+2M\frac{1-\beta}{\beta}\ln\left|\frac{U-4M\beta}{4M}\right|\,. \label{match} 
\ee 
We can verify that the Schwarzschild expression is reproduced for 
$\beta=1$. 

To ensure regularity of the modes, we require that they vanish at $r = 0 $ such that the origin acts like a moving mirror in the $(U,V)$ coordinates. Since there is no field behind $r<0$, the form of field modes can be determined, such that a $U\leftrightarrow v$ identification is made for the outgoing, Doppler-shifted right-movers \cite{Birrell:1982ix}. We are now ready to analyze the analog mirror trajectory by making the identification $u(U) \leftrightarrow f(v)$, a known function of the advanced time $v$. 

Using the standard moving mirror formalism \cite{Birrell:1982ix}, we study the massless scalar field in $(1+1)$-dimensional Minkowski spacetime (following e.g.\  \cite{Good_2015BirthCry}). From Eq.~(\ref{match}) the Kerr analog moving mirror trajectory is 
\be 
f(v)=v-\frac{1}{2g}\frac{1+\beta}{\beta}\ln|gv|+\frac{1}{2g}\frac{1-\beta}{\beta}\ln|gv-\beta|\,. 
\label{f(v)}\ee 
Note that the prefactor of the first logarithm is simply $1/\kappa$ 
(the surface gravity at $r_p$) and the prefactor of the second 
logarithm is $1/\kappa_m$ (the surface gravity at $r_m$) and so for $\beta=1$ the reduction to the Schwarzschild case (where 
$g=\kappa$) is clear. This is now to  be regarded as the trajectory of 
a perfectly reflecting boundary in flat spacetime rather than 
the origin as a function of coordinates in curved Kerr spacetime. We have reintroduced $g \equiv 1/(4M)$ to signal that we are now working in the moving mirror model with a background of flat spacetime, where $g$ is related to the acceleration parameter of the trajectory.

The rapidity, in advanced time, can be obtained via 
$-2\eta(v) = \ln f'(v)$, where the prime denotes a derivative with respect to the argument \cite{Good:2017ddq}. This yields
\be 
\eta(v)=-\frac{1}{2}\ln\left|1-\frac{1+\beta}{2\beta gv}+\frac{1-\beta}{2\beta(gv-\beta)}\right|\,. \label{rapidity} 
\ee 
The rapidity asymptotes at $ v = 0$, i.e.\ the mirror approaches the speed of light at the horizon, $u\rightarrow +\infty$. 
Again, this agrees with the Schwarzschild result for $\beta=1$.  
The proper acceleration $\alpha = e^{\eta(v)}\eta'(v)$ is also easily found and diverges as $v\rightarrow 0^-$, with  
$\alpha(v\to 0^-)=-\kappa/\sqrt{-4\kappa v}$. 
Thus, the late-time acceleration is related to the surface gravity $\kappa$, and both will be related to the temperature of the 
thermal spectrum of particles produced. The acceleration goes to zero at $v=-\infty$ as $-\beta/(g^2v^2)$.  

\begin{figure}[h]
    \centering
    \includegraphics[width=3in]{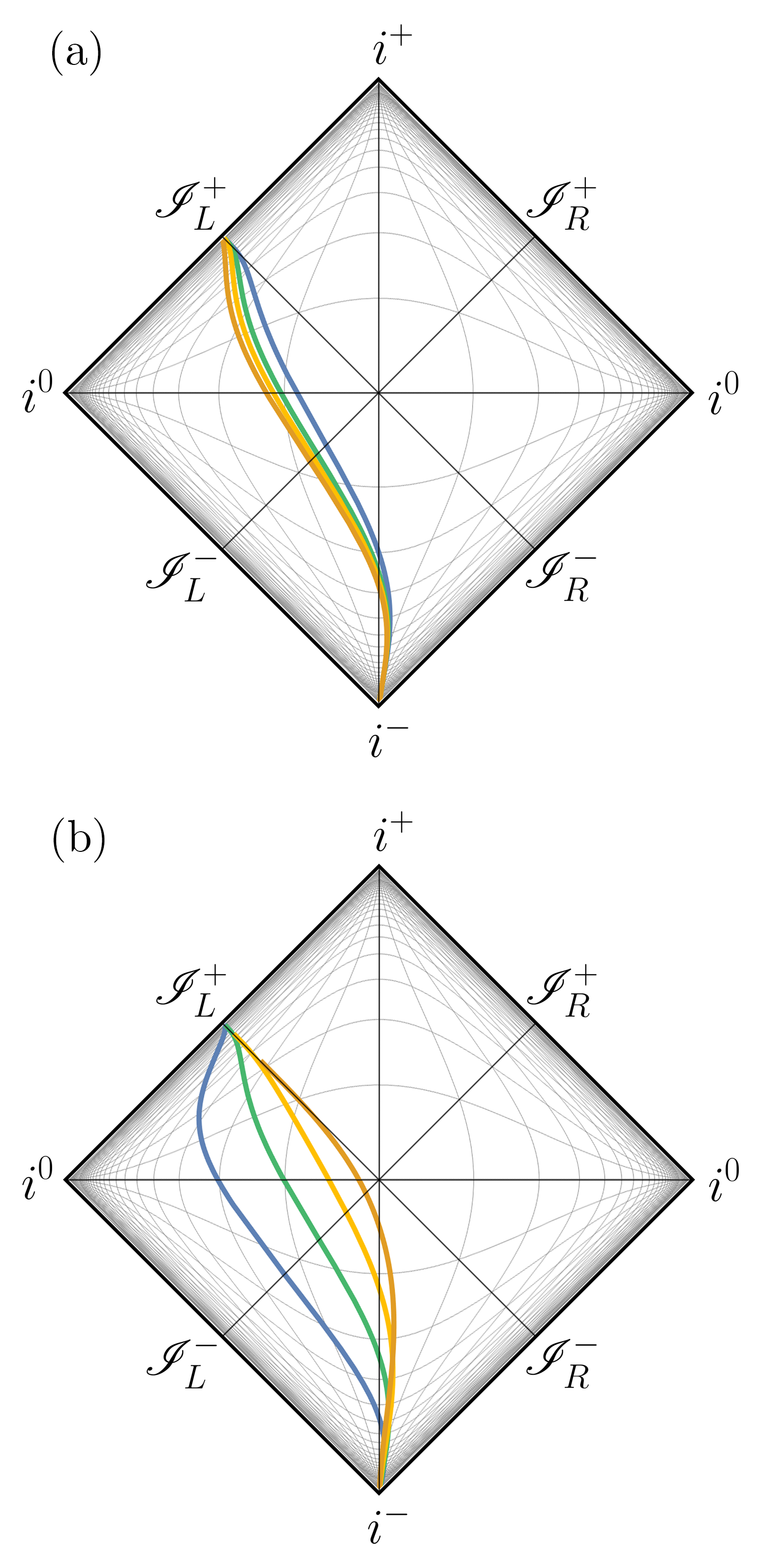}
    \caption{Penrose conformal diagrams for a class of generalized trajectories of the form Eq.~(\ref{f(v)}), with $v_H=0$. In (a), $g =1$ with
    $a/M= 0, 0.9, 0.975 , 0.999$ 
from blue to dark orange respectively. In (b), $\beta= 0.5$ ($a/M=0.866$) with $g=1/(4M)= 0.5, 1, 2, 5$, again from blue to dark orange. Notice the approach to the advanced time horizon at $v_H=0$, with the mirror asymptotically light-like.}
\label{fig:spacetimeplotbeta}
\end{figure}

The trajectory in a conformal Penrose diagram is plotted in Fig.~\ref{fig:spacetimeplotbeta}(a), showing the influence of the spin for different values of $\beta=\sqrt{1-a^2/M^2}$, while Fig.~\ref{fig:spacetimeplotbeta}(b) displays the 
influence of the mass for different values of $g=1/(4M)$.

\section{Flux, Spectrum, and Particles} \label{sec:particles} 
For the analog Kerr mirror, we find that the energy flux is constant at late times. The radiated energy flux $F(v)$ as computed from the quantum stress tensor can be calculated via the Schwarzian derivative of Eq.~(\ref{f(v)}) \cite{Good_2017Horizonless}, 
\be F(v)= \frac{1}{24\pi}\{f(v),v\}\,f'(v)^{-2}\,,\label{FLUX1}\ee
where the Schwarzian brackets are defined as
\be \{f(v),v\}\equiv \frac{f'''}{f'} - \frac{3}{2}\left(\frac{f''}{f'}\right)^2\,.\ee 
This yields, to leading order in $v$, near $v\rightarrow 0^-$,
\be F = \frac{\kappa^2}{48\pi}+\mathcal{O}(v^2)\,, \label{constantflux}\ee 
where $\kappa = g-k$. Here the spring constant arising from the angular momentum is 
$k=g(1-\beta)/(1+\beta)$, which 
vanishes for the Schwarzschild case and 
is maximal for the extreme Kerr case (forcing $\kappa$ to zero). 
See Appendix~\ref{appx} for more details. 

Equation~(\ref{constantflux}) is indicative of late time thermal equilibrium. 
From Fig.~\ref{fig:spacetimeplotbeta} we 
see that as $\beta$ decreases ($a/M$ increases), the late-time trajectory is further from the light-like asymptote (less accelerated), and so as expected, the particle flux decreases as does the temperature. We next present the derivation of the accompanying Planck distribution of the late time thermal equilibrium.

The particle spectrum can be obtained from the beta Bogoliubov coefficient, which can be found via \cite{Good_2017Horizonless} 
\be \beta_{\omega\omega'} = -\frac{1}{4\pi\sqrt{\omega\omega'}}\int_{-\infty}^{v_H} \d v ~e^{-i \omega' v -i \omega f(v)}\left(\omega f'(v)-\omega'\right)\,,\label{betaint}\ee 
where $\omega$ and $\omega'$ are the frequencies of the outgoing and incoming modes respectively. After integration by parts and neglecting the non-contributing surface terms, Eq.\ (\ref{betaint}) can be written 
\be \beta_{\omega\omega'} = \frac{1}{2\pi}\sqrt{\frac{\omega'}{\omega}}\int_{-\infty}^{v_H} \d v\: e^{-i\omega'v-i\omega f(v)}\,.\label{partsint}\ee
To obtain the particle spectrum, we take the modulus square, 
\be N_{\omega \omega'} \equiv |\beta_{\omega\omega'}|^2\,. \label{Kthermal}\ee
which gives 
\be N_{\omega \omega'} = \frac{\omega'}{2 \pi  \kappa \omega_p^2}\,\frac{e^{\pi(\omega/\kappa)(1-\beta)/(1+\beta)}}{e^{2 \pi  \omega /\kappa }-1}\,|U|^2\,,\label{up}
\ee 
where $U$ is the confluent hypergeometric Kummer function of the second kind, given by  
\be U\equiv U\left( \frac{i \omega}{\kappa}\frac{1-\beta}{1+\beta}, -\frac{i \omega }{g},-\frac{i \beta \omega_p}{g}\right)\,. 
\ee 
and $\omega_p = \omega'+\omega$.

\begin{figure}[ht]
\centering 
\includegraphics[width=\columnwidth]{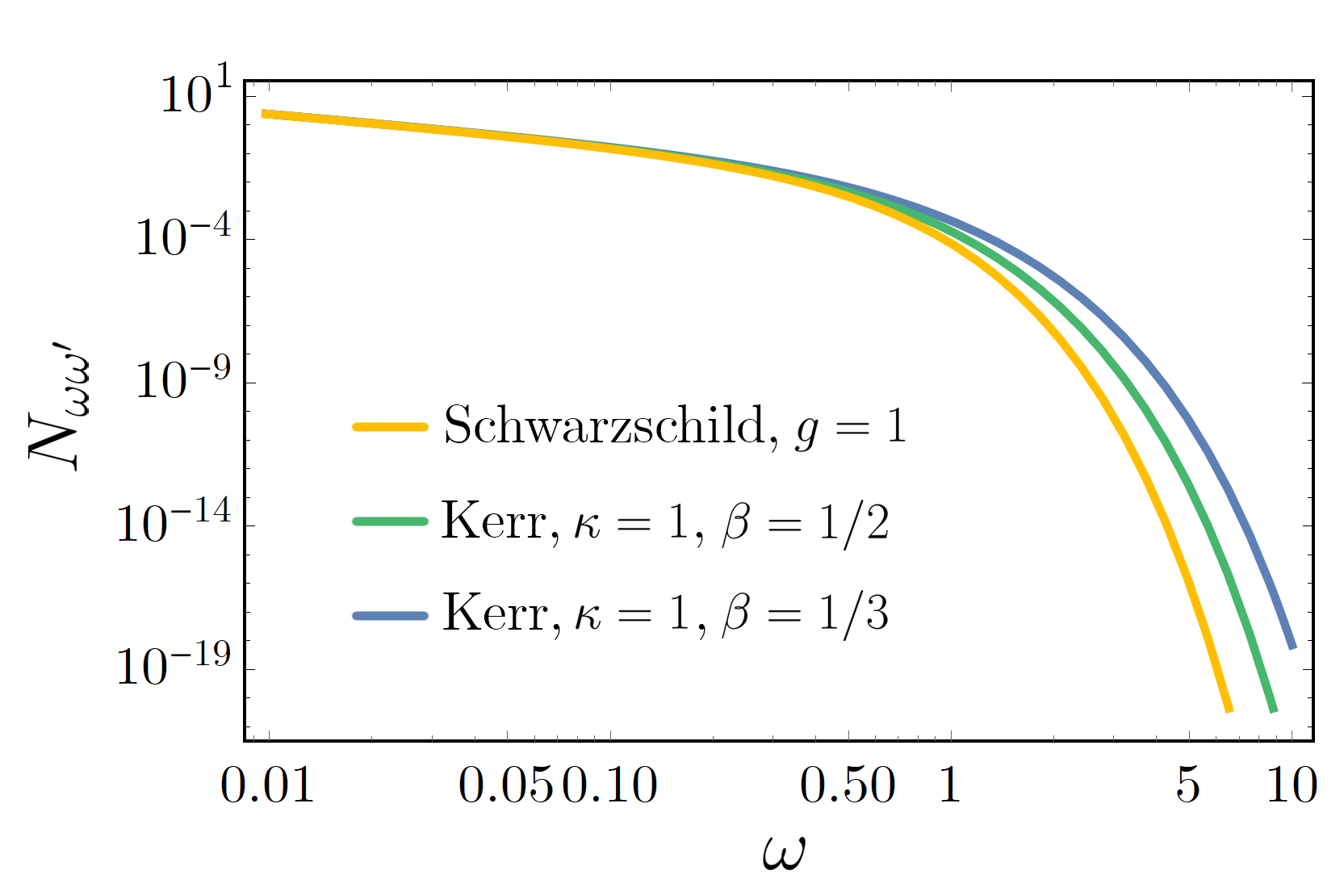} 
\caption{The mode spectra $N_{\omega\omega'} \equiv |\beta_{\omega\omega'}|^2$, setting $\omega'=1$ for illustration. At late times (i.e.\ $\omega\ll\omega'$) all approach thermal spectra. Here we have chosen the temperatures to be the same by setting $\kappa = 1$ for Kerr and $g=1/(4M)=1$ for Schwarzschild. 
}\label{fig3} 
\end{figure}

\begin{figure}[ht]
\centering 
\includegraphics[width=\columnwidth]{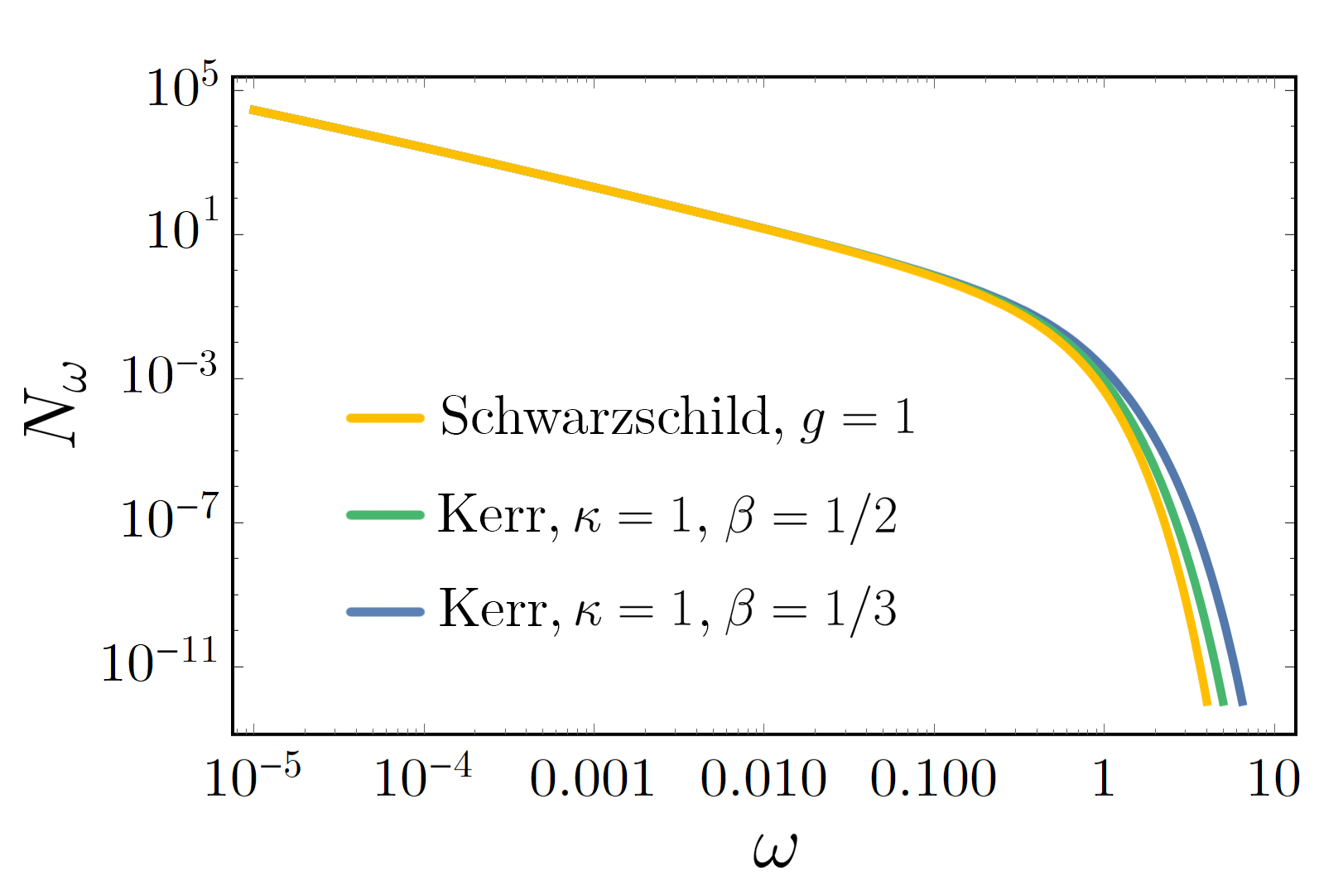} 
\caption{The particle count spectra $N_{\omega} =\int N_{\omega\omega'}d\omega'$. Both Kerr and Schwarzschild solutions have thermal spectra at temperature $\kappa/2\pi$ and $g/2\pi$ respectively. Here we have chosen $\kappa = 1$ for Kerr and $g=1\,(=\kappa)$ for Schwarzschild, with the same cases as Fig.~\ref{fig3}. }
\label{fig4} 
\end{figure}

The mode-mode spectrum $N_{\omega\omega'}$ is plotted in Fig.~\ref{fig3}.  
The particle spectrum $N_\omega$,
\be N_\omega = \int_0^\infty \mathrm{d}\omega' N_{\omega \omega'} \, \label{N(w)}\ee
is obtained numerically and plotted in Fig.~\ref{fig4}, illustrating a thermal Planck particle number spectrum at late times. Multiplying by the energy and phase space factors gives the usual Planck blackbody energy spectrum.   

Thermal behavior can also be seen analytically by a series approximation on $N_{\omega\omega'}$ in the high frequency limit $\omega'\gg \omega$, which amounts to late times, as first introduced by Hawking \cite{Hawking:1974sw}. As noted already, $\omega'$ is the frequency of the modes that have incoming, left-moving plane wave form, 
while $\omega$ is the frequency of the second set of modes that have outgoing right-moving plane wave form 
(see e.g. \cite{Good_2017Horizonless}). 
Late-time incoming modes become extremely red-shifted by the receding trajectory of the mirror. The main contribution to the beta Bogoliubov coefficient comes from these high-frequency incoming modes. Therefore they are governed by the asymptotic form for high-frequency,  which is independent of the details of collapse. Indeed in this limit (formally both $\omega'\gg\omega$ and $\beta\omega'\gg g$), the expressions only depend on the asymptotic acceleration, which is a function of $\kappa$ only, and not $g$ and $\beta$ separately. 

The result is
\be N_{\omega\omega'}^{K} = \frac{1}{2 \pi \kappa \omega'} \frac{1}{e^{2 \pi  \omega/\kappa }-1}\,, \quad \omega'\gg\omega\,, \label{eq:Kerrhi} 
\ee 
showing a late time equilibrium temperature of 
$T=\kappa/(2\pi)$. 

We can compare this to the Schwarzschild mirror \cite{Good_2016}, which has beta coefficient squared given by 
\be N^{S}_{\omega\omega'} :=|\beta_{\omega\omega'}^{\textrm{S}}|^2= \frac{\omega '}{2 \pi g \left(e^{2\pi \omega/g }-1\right) \left(\omega '+\omega \right)^2}\,,\label{Schw}\ee 
with $g=1/(4M)$ as usual. 
The Schwarzschild case corresponds to $\beta=1$, giving $\kappa=g$ by Eq.~(\ref{eq:tk}) or (\ref{eq:kappa}), and Eq.~(\ref{up}) reduces exactly to Eq.~(\ref{Schw}) by the identity $U(0,b,z)=1$ of the Kummer function. 

In the high frequency regime $\omega'\gg \omega$, where the incoming modes are extremely red-shifted, one has the per mode squared spectrum $N_{\omega \omega'} :=|\beta_{\omega\omega'}|^2$ as 
\be N^{S}_{\omega \omega'} = \frac{1 }{2\pi g \omega'}\frac{1}{e^{2\pi \omega/g }-1}\,,\quad \omega'\gg\omega ,\label{CWthermal}\ee 
with the Schwarzschild temperature given by $T=g/(2\pi)$.  Equation~(\ref{CWthermal}) is the eternal thermal spectrum for the mirror trajectory \cite{Good:2012cp} of Carlitz and Willey \cite{carlitz1987reflections}. 
As expected, Eq.~(\ref{eq:Kerrhi}) 
reduces to Eq.~(\ref{CWthermal}) for $\beta=1$ (i.e.\ $\kappa=g$).


\section{From Kerr to Extremal Kerr}\label{sec:EK} 

\subsection{Extremal Kerr} 

The extremal Kerr (EK) limit is defined by $J = M^2$, or $a=M$, corresponding 
to $\beta=0$. This means that $r_p=r_m=M$ and we have to redo our derivation in the 
$\beta=0$ limit to avoid division by zero in the last term of Eq.~(\ref{eq:tort}). The relevant radial and time pieces of the metric are given by
\be \d s^2 = -f(r)\d t^2 + f(r)^{-1}\d r^2,  \ee with 
\be f(r)= 1 - \frac{2M r}{r^2 + M^2}\,.\ee 
giving a horizon at $r=M$. The tortoise coordinate is found by the usual integration Eq.~(\ref{tort}), giving
\be r^* = r-M+ \frac{2 M^2}{M-r}+2 M \ln \frac{M-r}{2M}\,.\ee 
We perform the standard analysis by matching $r^* = r$, solving for the trajectory of the center and then applying the regularity condition of the modes. This gives the moving mirror trajectory, 
\be f(v) = v+2M-\frac{8 M^2}{v}-4 M \ln \left(-\frac{v}{4M}\right)\,.\label{f(v)EK}\ee
Here we have set the shell to $v_0 = 2M$, so that the horizon is at $v_H=0$. Note that the constant term $2M$ in $f(v)$ has no effect on 
the acceleration or flux. 

To signal that we are now in flat space with an accelerated boundary, we associate $M$ with $\mathcal{A}$, where $\mathcal{A}$ is the limiting uniform proper acceleration of the mirror at late times, 
\be \lim_{v\rightarrow 0} \alpha(v) = -\mathcal{A}\,.\ee
This gives $M = 1/(2\sqrt{2}\mathcal{A})$. Thus the extremal Kerr case gives asymptotic 
uniform acceleration, and so the energy flux vanishes in  this limit $v\to0$. This is behavior in common with asymptotically inertial mirrors. The model still produces an infinite total particle count in contrast to asymptotic zero velocity (static) mirrors like that proposed by Walker and Davies \cite{Walker_1982}, or the `Schwarzschild mirror with quantum purity' model \cite{Good:2019tnf,GoodMPLA, good2020schwarzschild}, which yield finite total particle count $N$.
An infinite total particle count is expected from the extremal Kerr case ($\beta=0$) because infinite soft particles (zero frequency) -- an IR divergence -- is present for uniform acceleration.

\begin{figure}[bth]
\centering 
\includegraphics[width=\columnwidth]{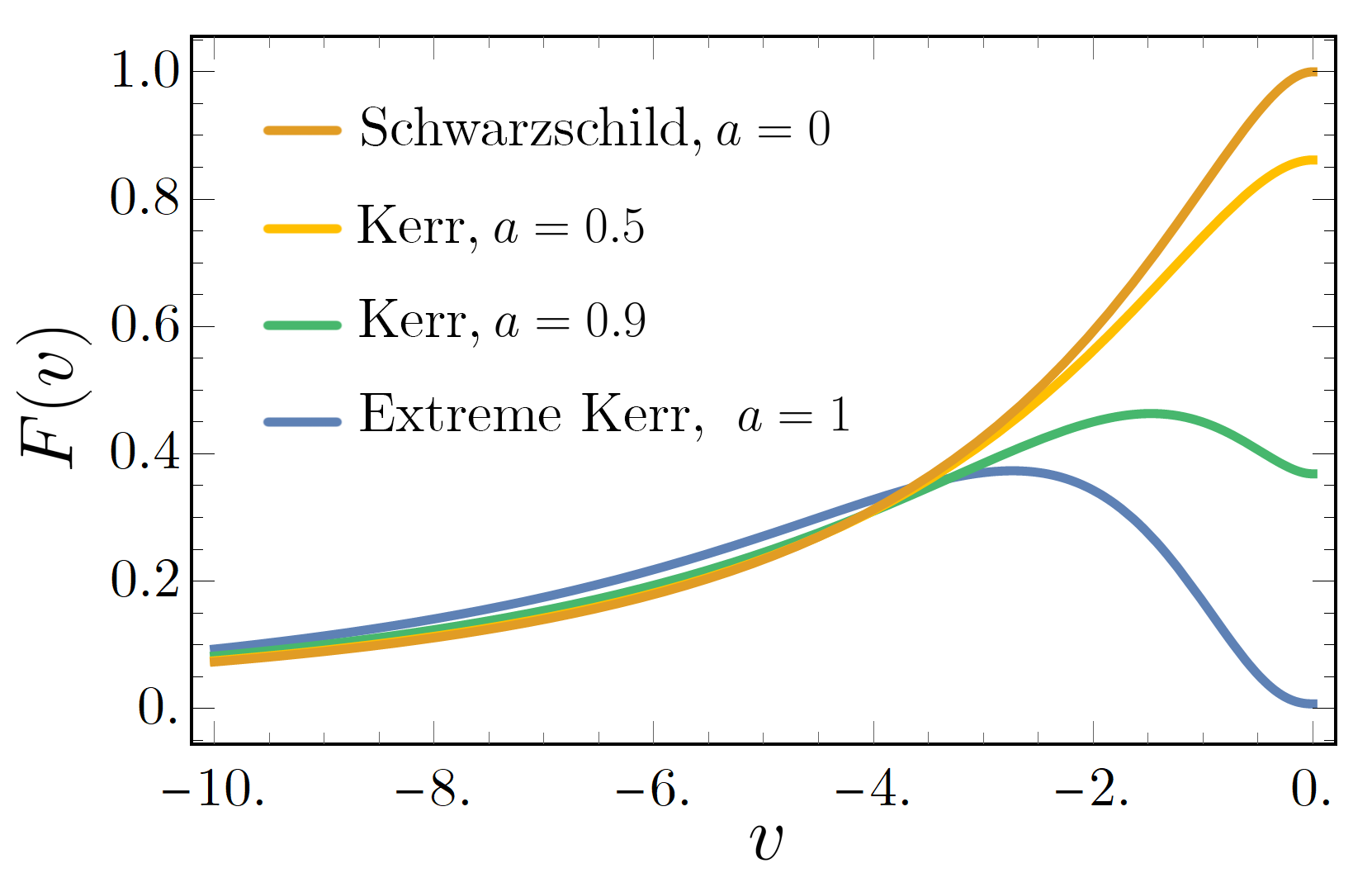} 
\caption{The energy flux as computed by the stress tensor. The total energy from the non-extremal stress tensors are infinite because the energy flux is eternally thermal and observation happens over retarded time: $E = \int F du$.  The results have been normalized by $768\pi$ (i.e.\ $M=1$) so that maximum Schwarzschild thermal emission is $F(v=0)= 1$.
}
\label{fig:Energy} 
\end{figure}

The energy flux is straightforward to derive from Eq.~(\ref{FLUX1}), 
\be F(v) = -\frac{M v^3 \left(8 M^2-7 M v+v^2\right)}{3 \pi  \left(8 M^2-4 M v+v^2\right)^4}\,,\label{stressEK}\ee
and is plotted as the blue curve in 
Fig.~\ref{fig:Energy}. 

The total stress energy is found via substitution of Eq.~(\ref{f(v)EK}) and Eq.~(\ref{stressEK}) into
\be E = \int_{-\infty}^{v_H=0} F(v) \frac{\mathrm{d} f(v)}{\mathrm{d}v} \mathrm{d}v\,, \ee
which gives
\be E = \frac{\mathcal{A}}{48\pi}\frac{\pi -1}{ \sqrt{2}}\,.\label{stresstotenergyEK}\ee 
The particle mode-mode spectrum via substitution of Eq.~(\ref{f(v)EK}) into Eq.~(\ref{betaint}) or Eq.~(\ref{partsint}) gives,
\be |\beta_{\omega\omega'}|^2 = \frac{e^{-\sqrt{2} \pi  \omega/\mathcal{A} } \,\omega ' }{\pi ^2 \mathcal{A} ^2 \omega _p}\left|K_j\left(\frac{2}{\mathcal{A} }\sqrt{\omega  \omega _p}\right)\right|^2\,,\label{betasqEK}\ee
where $\omega_p \equiv \omega + \omega'$, and $j \equiv 1+i\omega\sqrt{2}/\mathcal{A}$. The total energy carried by the particles is the same as that derived by the stress tensor radiation, Eq.~(\ref{stresstotenergyEK}),
\be E = \int_0^{\infty}  \mathrm{d}\omega\int_0^\infty \mathrm{d}\omega' \omega  |\beta_{\omega\omega'}|^2 = \frac{\mathcal{A}}{48\pi}\frac{\pi -1}{ \sqrt{2}}\,.\ee

\subsection{Comparison with Extreme Reissner-Nordstr\"om} 

\begin{figure}[ht]
\centering 
\includegraphics[width=\columnwidth]{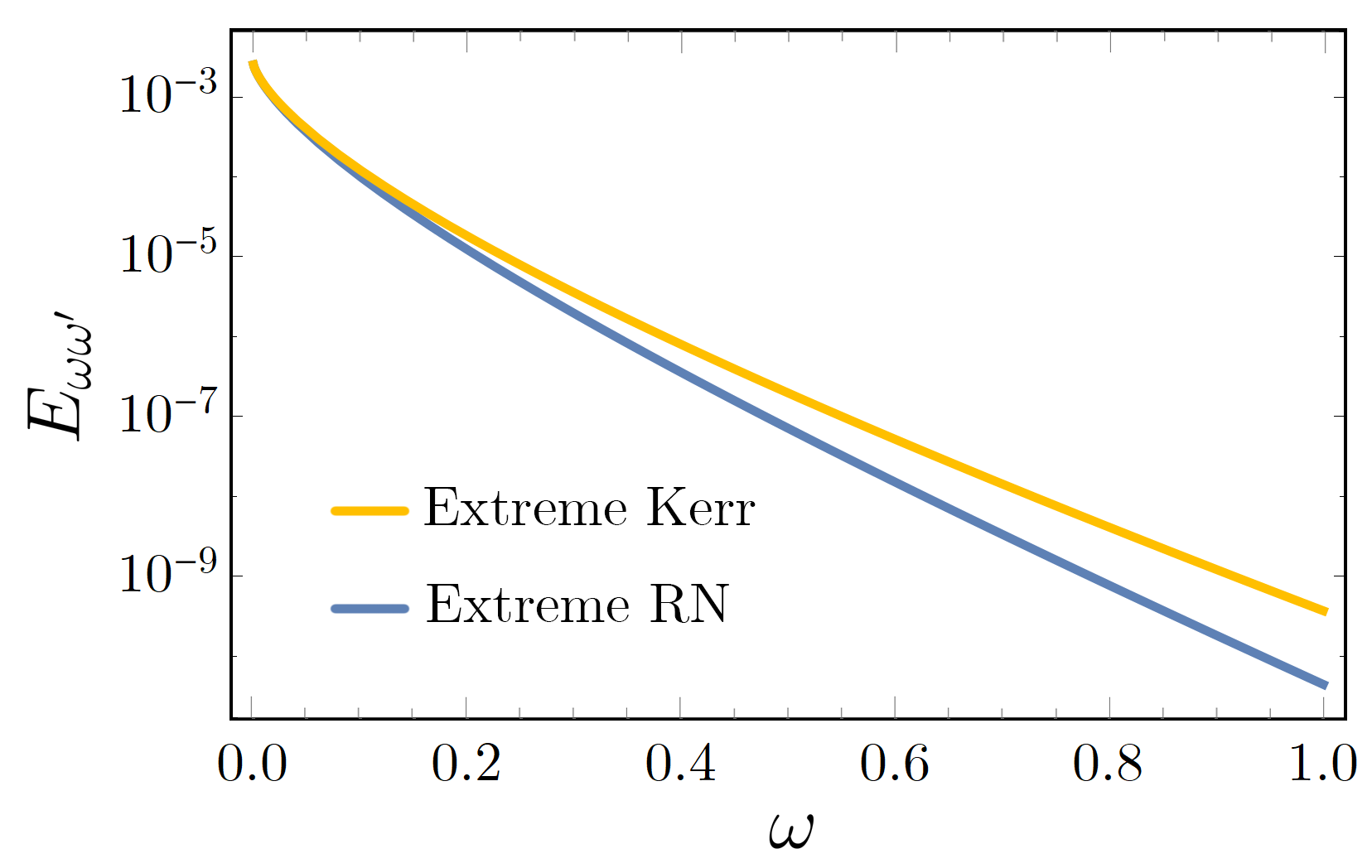} 
\caption{The integrand $E_{\omega\omega'}=\omega|\beta_{\omega\omega'}|^2$ for extremal Kerr and extremal RN with equal asymptotic uniform accelerations, $\mathcal{A} = 1$. Here we have set $\omega' = 10$.  The surface gravities of the extreme black holes are zero, i.e.\ the temperature is undefined as at no point during collapse do the particles find themselves distributed according to a Planck spectrum. 
}
\label{fig:EKvsERN} 
\end{figure}

\begin{figure}[ht]
\centering 
\includegraphics[width=\columnwidth]{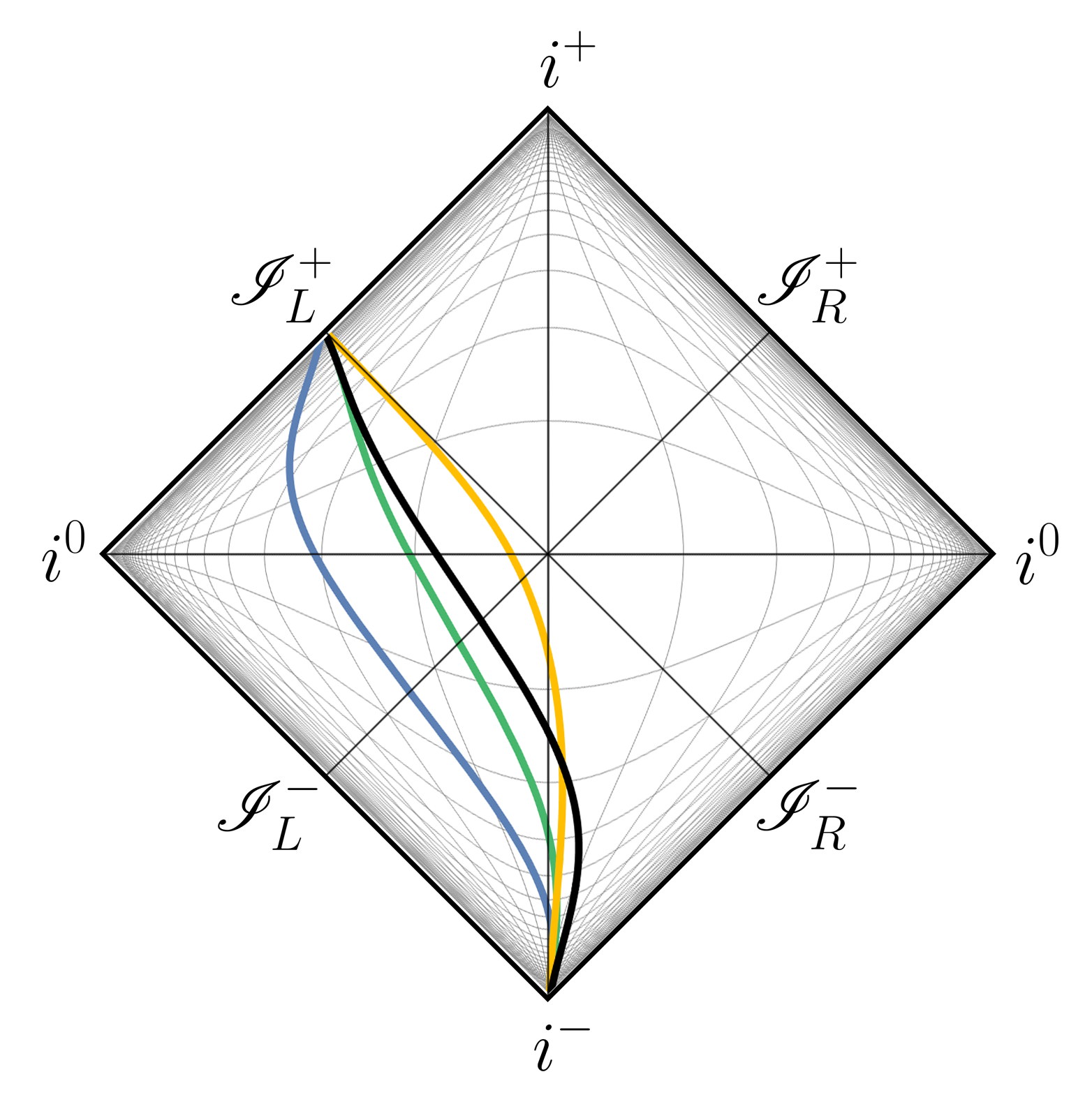} 
\caption{Conformal Penrose diagram for extremal mirror trajectories. The extremal Kerr mirrors are the blue to yellow trajectories with asymptotic uniform acceleration $\mathcal{A} = 1, 2, 8$, respectively. The extremal RN trajectory \cite{good2020extreme} is in black, with $\mathcal{A}=2$.
}
\label{fig:EKpenrose} 
\end{figure}

We can compare Eq.~(\ref{betasqEK}) for the extreme Kerr (EK) case 
with the extreme Reissner-Nordstr\"om (ERN) case \cite{good2020extreme}, 
\be |\beta_{\omega\omega'}|^2_\text{ERN}=\frac{e^{-2 \pi  \omega /\mathcal{A}_\text{ERN} } \omega ' }{\pi ^2 \mathcal{A}_\text{ERN} ^2 \omega _p}\left|K_q\left(\frac{2}{\mathcal{A}_\text{ERN} }\sqrt{\omega  \omega _p}\right)\right|^2\,,\label{ERNbeta2}\ee
where $q \equiv 1+2i \omega/\mathcal{A}_{\rm ERN}$. The forms are  
quite similar but differ in the details. 
For two equal mass ERN and EK black holes, the ERN emits more total energy than the EK,   
\bea 
E_\text{ERN} &=& \frac{1}{72\pi M_\text{ERN}} \simeq  \frac{0.0044}{M_\text{ERN}} \\ 
E_\text{EK}&=& \frac{\pi -1}{192 \pi  M_\text{EK}} \simeq \frac{0.0035}{M_\text{EK}}\,.
\eea 
Of course, the inverse relationship between acceleration and mass reverses the situation if instead we take two equal late-time acceleration ERN and EK mirrors. Then the EK emits more total energy than the ERN. Since  $\mathcal{A}_\text{ERN} = 1/(2M_\text{ERN})$ and $\mathcal{A}_\text{EK} = 1/(2\sqrt{2}M_\text{EK})$, 
we have 
\begin{align}
    E_\text{ERN} &= \frac{\mathcal{A_{\rm ERN}}}{36\pi} \simeq 0.009 \mathcal{A_{\rm ERN}} \\
    E_\text{EK} &= \frac{\mathcal{A}_\text{EK}}{48\pi} \frac{\pi - 1}{\sqrt{2}} \simeq 0.010 \mathcal{A_{\rm EK}}\,. 
\end{align}
Figure~\ref{fig:EKvsERN} compares the integrand of $E$ for the two extremal mirrors, $\omega |\beta_{\omega\omega'}|^2$. Figure~\ref{fig:EKpenrose} shows the Penrose diagram of EK for various asymptotic accelerations, with a comparison to an ERN trajectory in black.

\section{Conclusions}\label{sec:conc} 

In this paper, we have derived the particle spectrum for a rotating Kerr black hole by utilizing the well-known accelerating boundary correspondence in (1+1)-dimensional flat spacetime. 
Kerr black holes are particularly interesting to understand since they represent the observed black holes in our universe, and the accelerating mirror 
approach makes the calculations tractable. Solving for the beta Bogoliubov coefficients, we find they can be written in terms of special functions and give rise to a particle spectrum with a late-time Planck distribution with temperature proportional to the surface gravity, $\kappa = g-k$. We have seen that the angular coordinates are degenerate and irrelevant for computing the correct temperature at late times, reminiscent of the holographic principle and the 1-dimensional nature of information flow \cite{Padmanabhan:2002sha} for black holes \cite{Bekenstein:2001tj}. 

We presented results for the continuous range from 
Schwarzschild ($\beta=1$) to extreme Kerr ($\beta=0$) solutions, also comparing to  
the extreme Reissner-Nordstr{\"o}m case. 
In particular the temperature of the late-time Planck spectrum decreases from the Schwarzschild case as one approaches maximal spin, where the flux finally vanishes.

The accelerated boundary correspondence continues to be demonstrated as a useful tool, here enabling us to use our derived Kerr moving mirror solution to confirm that the distribution of particles produced from a Kerr spacetime at late times is the thermal Planck spectrum, with temperature related to the surface gravity, or alternately mirror acceleration. As mentioned in the Introduction, several notable geometries, including the Schwarzschild, Reissner-Nordstr\"om, and the de Sitter/anti-de Sitter spacetimes have recently been studied. The utility of this approach should allow for its application to more complex metrics such as asymptotically de Sitter/anti-de Sitter black holes, and accelerated black holes described by the $C$-metric \cite{griffiths2009exact}.

\acknowledgments 

Funding from state-targeted program ``Center of Excellence for Fundamental and Applied Physics" (BR05236454) by the Ministry of Education and Science of the Republic of Kazakhstan is acknowledged. M.G.\ is also funded by the FY2018-SGP-1-STMM Faculty Development Competitive Research Grant No. 090118FD5350 at Nazarbayev University. J.F.\ acknowledges support from the Australian Research Council Centre of Excellence for Quantum Computation and Communication Technology (Project
No.\ CE170100012). E.L.\ is supported in part by the Energetic Cosmos Laboratory and by the U.S.\ Department of Energy, Office of Science, Office of High Energy Physics, under Award DE-SC-0007867 and contract no.\ DE-AC02-05CH11231.

\appendix
\section{Derivation of $\kappa = g-k$} \label{appx}
The first law of black hole mechanics relates the two essential parameters, $(M,J)$, the mass and angular momentum of a rotating black hole:
 \begin{equation}
 dM = \frac{\kappa}{8\pi} dA + \Omega \,dJ \,, 
 \end{equation}
 where $A$ is the outer horizon area, $\kappa$ is the outer surface gravity, $\Omega$ is the outer angular velocity, and $J$ is the angular momentum.  
 The area is known,
 \begin{equation}
 A = 4\pi (r_p^2 + a^2) = 4\pi r_s r_p \,, 
 \end{equation}
 where $r_p = M + \sqrt{M^2 - a^2}$, $a = J/M$ and $r_s = 2M$. Equivalently, 
 \begin{equation}
 M^2 = \frac{A}{16\pi} + 4\pi \frac{J^2}{A}\,.
 \end{equation}
 Therefore,  one can differentiate both sides to get,
 \begin{equation}
 \kappa = 8\pi \left.\frac{\partial M}{\partial A}\right|_J = \frac{1}{4M} - M \left(\frac{4\pi J}{M A}\right)^2\,.
 \end{equation}
 Since 
 \begin{equation}
 \Omega = \left.\frac{\partial M}{\partial J}\right|_A = \frac{4\pi J}{M A}\,,
 \end{equation}
 we have $ \kappa = g - k$ where $g = 1/(4M)$ and $k = M \Omega^2$.  
 
 This can also be obtained by the usual formula:
 \be \kappa = \frac{1}{2}\left. \frac{d}{dr}f(r)\right|_{r=r_p} = \frac{1}{2M} - \frac{1}{2r_p}= \frac{1}{4M} - \frac{a^2}{4M r_p^2}\,,\ee 
thus $\kappa = g - k$.
 
Recalling the notation $\beta=\sqrt{1-a^2/M^2}$, 
we see that $A=8\pi M^2(1+\beta)$ and 
 \be 
 k=M\left(\frac{4\pi J}{MA}\right)^2=\frac{1}{4M}\frac{1-\beta}{1+\beta}\,, 
 \ee 
 so 
 \be 
 \kappa=\frac{1}{4M}\frac{2\beta}{1+\beta}=
 \frac{2g\beta}{1+\beta}\,, \label{eq:kappa}
 \ee 
 having the correct Schwarzschild ($\beta=1$) and extremal Kerr ($\beta=0$) limits.


\bibliography{main}

\end{document}